# Fast readout algorithm for cylindrical beam position monitors providing good accuracy for particle bunches with large offsets

P. Thieberger*, D. Gassner, R. Hulsart, R. Michnoff, T. Miller, M. Minty and Z. Sorrell

Collider Accelerator Department, Brookhaven National Laboratory, Upton, NY 11973, USA

and A. Bartnik, CLASSE, Cornell University, Ithaca, New York 14850, USA

**Abstract**

A simple, analytically correct algorithm is developed for calculating "pencil" relativistic beam coordinates using the signals from an ideal cylindrical particle beam position monitor (BPM) with four pickup electrodes (PUEs) of infinitesimal widths. The algorithm is then applied to simulations of realistic BPMs with finite width PUEs. Surprisingly small deviations are found. Simple empirically determined correction terms reduce the deviations even further. The algorithm is then tested with simulations for non-relativistic beams. As an example of the data acquisition speed advantage, a FPGA-based BPM readout implementation of the new algorithm has been developed and characterized. Finally, the algorithm is tested with BPM data from the Cornell Preinjector.

## I - Introduction

Retrieving accurate position information from particle beam position monitors (BPMs) often requires iterative computations such as least squares fitting in two dimensions that can be time-consuming and may limit data acquisition rates. Accuracy may also be impacted if the number of iterations needs to be restricted or if fewer measurements must be used when averaging is important for noise reduction.

When BPMs with cylindrical symmetry are used to determine the position of pencil beams[1], simplifications become possible that can mitigate these limitations. These limitations are particularly significant when beam offsets with respect to the cylindrical BPM axis are a significant fraction of the cylinder radius, where the non-linearity of the signal response becomes large enough to be important.

A significant simplification was achieved by C. Gulliford et.al. [1]. They start with the expression[2] for wall current density $J_z(\theta)$ induced by an off-center pencil beam on the interior surface of the cylinder (which they re-derive):

---

[1] A pencil beam is defined here as a beam that propagates along a line parallel to the axis of the cylindrical BPM and which has a diameter that is negligibly small compared to the diameter of the cylinder. The usual approximation (see e.g. refs. 2 and 3, and references therein) of representing such beams by continuous line charges, and solving the electrostatic problem is made in ref. 1 and will also be used in the present work.

$$J_z(\theta) = \frac{I}{2\pi a} \times \frac{r^2 - a^2}{a^2 + r^2 - 2ar \cos(\theta)} \qquad 1)$$

where $a$ is the radius of the BPM, $r$ is the radial component of the beam position and $\theta$ is the angle between the planes defined by the beam and by the line along the cylinder surface where the current density is calculated. This equation is strictly valid only for highly relativistic beams such as encountered in electron accelerators, the Relativistic Heavy Ion Collider (RHIC) and the Large Hadron Collider (LHC). As discussed by R. E. Shafer [3], the deviations that occur for low-$\beta$ beams have impacts on BPM sensitivity and linearity, which, in addition to being $\beta$-dependent, are also dependent on bunch length, BPM geometry and signal processing frequencies.

Using equation 1), C. Gulliford et.al. [1] find an analytical expression for the integral over the angles subtended by four symmetrically located stripline pickup electrodes. Using this expression simplifies and accelerates the least squares calculation used to solve the inverse problem; namely finding the beam position that best reproduces the measured signals.

In the present work, we use Eq. 1 for the case of BPMs consisting of very narrow striplines or very small buttons to obtain an analytical solution for the beam position, thus solving the inverse problem without iterative fitting procedures. Next, we test this solution simulating signals from a button BPM with Particle Studio[4] and find modest deviations due to the fact that the buttons are flat and not particularly small. The relatively small deviations found are then further reduced by developing a simple empirical correction. Next, we analyze deviations as a function of stripline widths. Then we study examples of deviations that occur if the new algorithm is applied non-relativistic beams. Next, a proposed FPGA implementation of the new algorithm is presented and analyzed. Finally, we use actual BPM data to test the new algorithm.

Reducing what was hitherto a two-dimensional fitting problem to straightforward numeric calculations results in large gains in processing speed and latency reduction, which will be important where high-rate data acquisition is required.

## II- Derivation of the analytic solution

Figure 1 shows a schematic cross section of a cylindrical BPM of radius $a$, with four symmetrically located buttons or striplines which are narrow enough so that the signals will be nearly proportional to the values of the wall current densities calculated at their centers by using Eq. 1 . We omit the constant $\frac{I}{2\pi a}$ which is the same for the four pickup electrodes (PUEs) and we find values $A_x$ and $B_x$ proportional to the respective signal amplitudes:

$$A_x = \frac{r^2 - a^2}{a^2 + r^2 - 2ar \cos(\theta)} \qquad 2)$$

$$B_x = \frac{r^2 - a^2}{a^2 + r^2 - 2ar \cos(\theta - \pi)} = \frac{r^2 - a^2}{a^2 + r^2 + 2ar \cos(\theta)} \qquad 3)$$

Calling $\rho = r/a$ we rewrite 2) and 3)

$$A_x = \frac{\rho^2 - 1}{\rho^2 + 1 - 2\rho \cos(\theta)} \qquad \qquad 4)$$

$$B_x = \frac{\rho^2 - 1}{\rho^2 + 1 + 2\rho \cos(\theta)} \qquad \qquad 5)$$

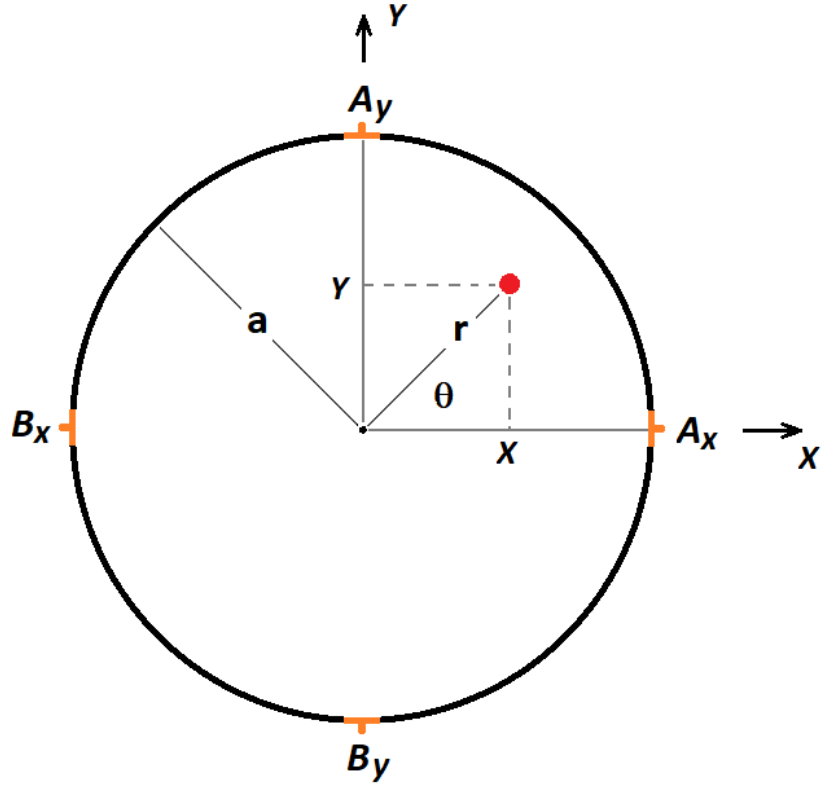

Figure 1 – Schematic of a cylindrical BPM with very small buttons or narrow striplines

Next, we write the usual $\frac{A_X - B_X}{A_X + B_X}$ ratios by using 4) and 5)

$$\frac{A_X - B_X}{A_X + B_X} = \frac{(\rho^2-1)(\rho^2+1+2\rho \cos(\theta)) - (\rho^2-1)(\rho^2+1-2\rho \cos(\theta))}{(\rho^2-1)(\rho^2+1+2\rho \cos(\theta)) + (\rho^2-1)(\rho^2+1-2\rho \cos(\theta))} \qquad 6)$$

$$\frac{A_X - B_X}{A_X + B_X} = \frac{4(\rho^3 - \rho)\cos(\theta)}{2(\rho^2-1)(\rho^2+1)} = \frac{2\rho \cos(\theta)}{\rho^2 + 1} \qquad \qquad 7)$$

Defining the values of $Q_x$ and $Q_y$ as:

$$Q_x = \frac{A_X - B_X}{A_X + B_X} \text{ and } Q_y = \frac{A_y - B_y}{A_y + B_y} \qquad \qquad 8)$$

we write:

$$Q_x = \frac{2\rho \cos(\theta)}{\rho^2+1} \quad \text{and} \quad Q_y = \frac{2\rho \sin(\theta)}{\rho^2+1} \qquad 9)$$

Where we have repeated the above derivation for the vertical plane, taking into account that $\cos(\theta - \pi/2) = \sin(\theta)$

We see that $Q_x$ and $Q_y$ are the components of a vector of modulus

$$Q = \sqrt{Q_x^2 + Q_y^2} = \frac{2\rho}{\rho^2+1} \qquad 10)$$

pointing in the direction of the X, Y beam position.

Rewriting 10:

$$Q\rho^2 - 2\rho + Q = 0 \qquad 11)$$

we get:

$$\rho = \frac{2 \pm \sqrt{4-4Q^2}}{2Q} = \frac{1 \pm \sqrt{1-Q^2}}{Q} = \frac{1}{Q} - \sqrt{\frac{1}{Q^2} - 1} \qquad 12)$$

where we had to choose the negative sign because $\rho = r/a$ must be smaller than *1* for the beam to be inside of the beam pipe. The positive sign solution corresponds to the position of the image charge.

Now using equations 9), 10) and 12) and remembering our definition $\rho = r/a$ where a is the radius of the BPM, we get the beam coordinates X and Y:

$$X = a\,\rho\,\frac{Q_x}{Q} \qquad 13)$$

$$Y = a\,\rho\,\frac{Q_y}{Q} \qquad 14)$$

since $\cos(\theta) = \frac{Q_x}{Q}$ and $\sin(\theta) = \frac{Q_y}{Q}$

For very small values of $Q$ which correspond to beam positions very close to the axis, the linear approximation is adequate and instead of 12) we use the first term of its Taylor expansion around $Q = 0$ which is[5] $Q/2$. Therefore, instead of 13) and 14) we can use:

$$X = a\,\frac{Q_x}{2} \qquad 15)$$

$$Y = a\,\frac{Q_y}{2} \qquad 16)$$

The paraxial approximation represented by equations 15) and 16) is in agreement with previous results such as equation 5.5 in an article by E. Shafer[6], when the limit is taken for infinitesimal PUEs. Having found these solutions, we will now verify them for a specific case in the next section.

**III - Verification with Particle Studio simulations of a small button BPM.**

The BPM used for the Particle Studio simulations is shown in Fig. 2

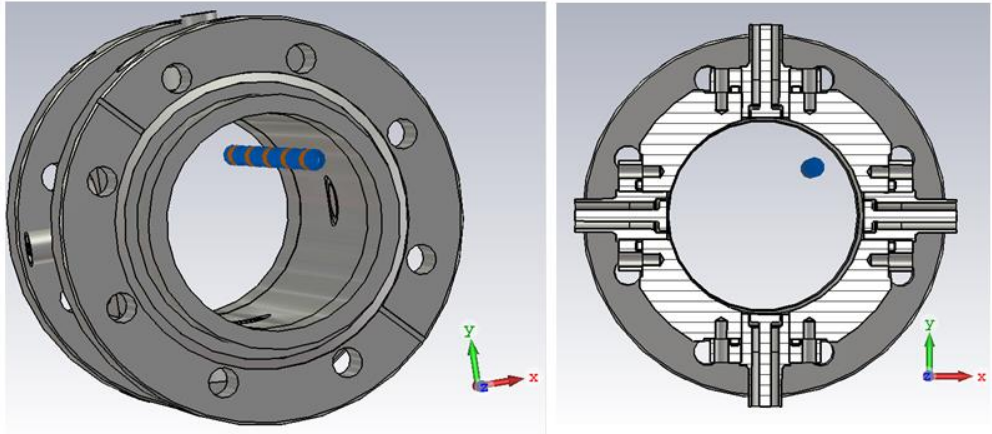

Figure 2 – Perspective view and cross-section of the BPM model used for the simulations. The BPM diameter is 2a = 60 mm and the button diameters are 10 mm. The beam position shown is X= 17.5 mm, Y= 17.5 mm.

Simulations were performed for beam positions from 0 to 20 mm in 5 mm steps in both dimensions. The assumed bunch charge was 1 nC and the Gaussian bunch length was 30 mm RMS. The simulations are performed for fully relativistic beams ($\beta$ = v/c = 1). The resulting signals from two opposite PUEs are shown in Fig. 3 for the case X=17.5, Y=17.5 mm.

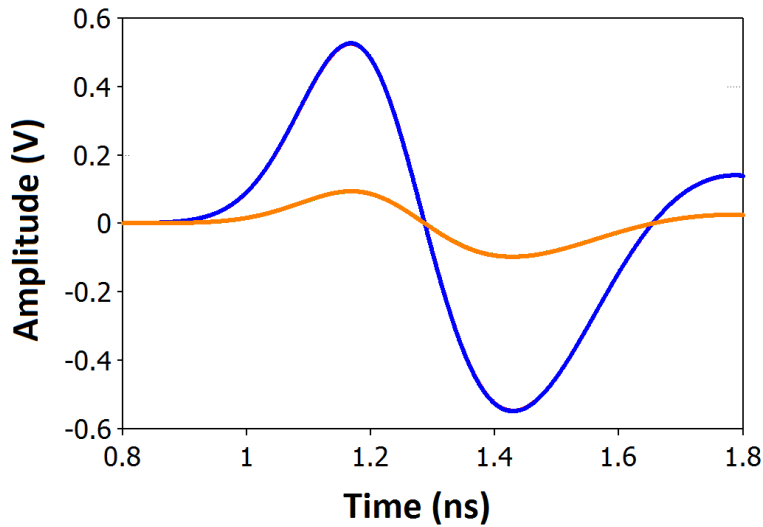

Figure 3 – Particle Studio outputs for opposite buttons using the BPM model shown in Fig. 2 for a beam located at X= 17.5 mm, Y= 17.5 mm.

The amplitudes of the simulated signals were then used to compute beam positions using the algorithm described in the previous section (equations 8, 10, 12, 13 and 14) . The results are shown in Fig. 4, where the red circles indicate the calculated positions while the black dots at the gridline intersections are the beam positions used in the simulations. The only adjustment that was made to improve the agreement was a 1.9% increase in the BPM diameter used in the calculations. This is due to the fact that the buttons, due to their flat surfaces, are partially recessed. Fig. 5 shows the distances between "real" and calculated positions along the X and Y axes and along the diagonal. These relatively small deviations are attributed to the size and shape of the buttons. Particle Studio simulation inaccuracies would contribute to these deviations too.

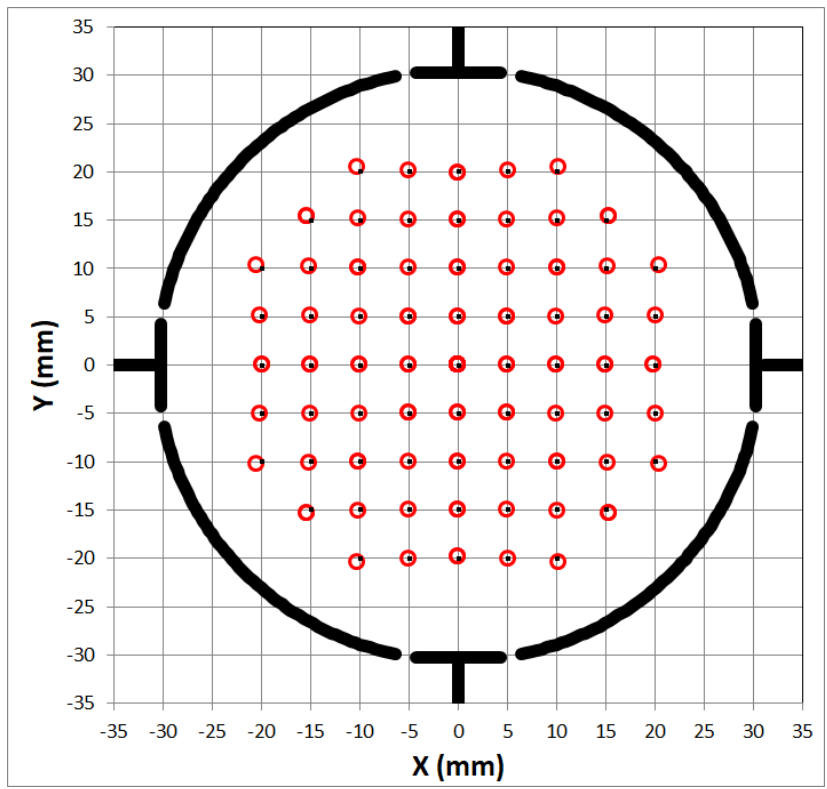

Figure 4 - Simulation results for a 60 mm diameter BPM showing position errors when using a two-dimensional analytical solution that is only strictly valid for infinitely small buttons. A 1.9% correction was made to the diameter. The observed deviations are due to the fact that the buttons have a 10 mm diameter and are flat instead of following the cylindrical contour of the BPM chamber. The RMS distance between calculated (circles) and nominal positions (dots) is 230 μm.

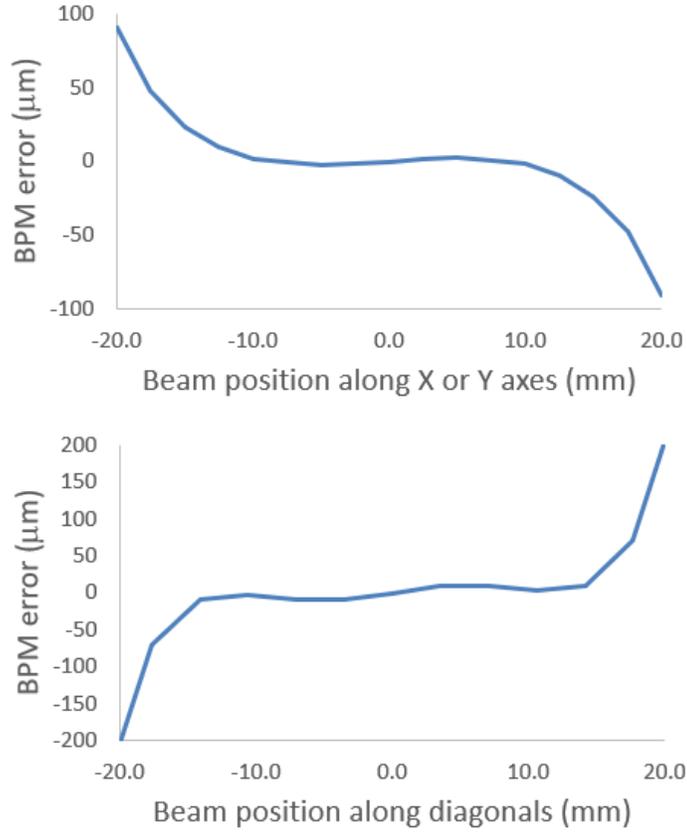

Figure 5 - Distances between the beam positions used in the simulations and positions calculated with the algorithm developed here using the simulation outputs plotted along the X and Y axes and along the diagonal.

## IV - Further empirical refinement

The analytic approach presented in the previous section is only strictly valid for perfect cylindrical symmetry and for infinitely small buttons or line-like striplines. We have seen that results from simulations for 10 mm diameter buttons in a 30-mm radius chamber are reproduced quite well by the simple algorithm after a minor adjustment of the diameter used in the calculations. The RMS distance between the calculated and simulated positions for the beam positions shown in Fig. 4 is 230 µm.

Noting that the largest deviations occur along the diagonals, we introduce an empirical correction factor that modifies the $Q_x$ and $Q_y$ values defined in section **II** for beam positions distant from the X=0 and Y=0 planes. Starting with equations 8),

we define corrected values $Q'_x$ and $Q'_y$ :

$$Q'_x = Q_x + b\, Q_x\, |Q_y| \qquad 17)$$

$$Q'_y = Q_y + b\,|Q_x|\,Q_y \qquad 18)$$

Where b is an adjustable parameter.

We then proceed as before, using now the primed quantities:

$$Q' = \sqrt{Q'^2_x + Q'^2_y} \qquad 19)$$

$$\rho' = \frac{1}{Q'} - \sqrt{\frac{1}{Q'^2} - 1} \qquad 20)$$

And finally:

$$X' = a(1+\epsilon)\,\rho'\,\frac{Q'_x}{Q'} \qquad 21)$$

$$Y' = a(1+\epsilon)\,\rho'\,\frac{Q'_y}{Q'} \qquad 22)$$

The parameter $\epsilon$ represents the small adjustment to the value of the radius we had mentioned in the previous section. Computationally these corrections add little additional time. The results of optimizing **b** as well as $\epsilon$ for the present example are shown in Figs. 6 and 7.

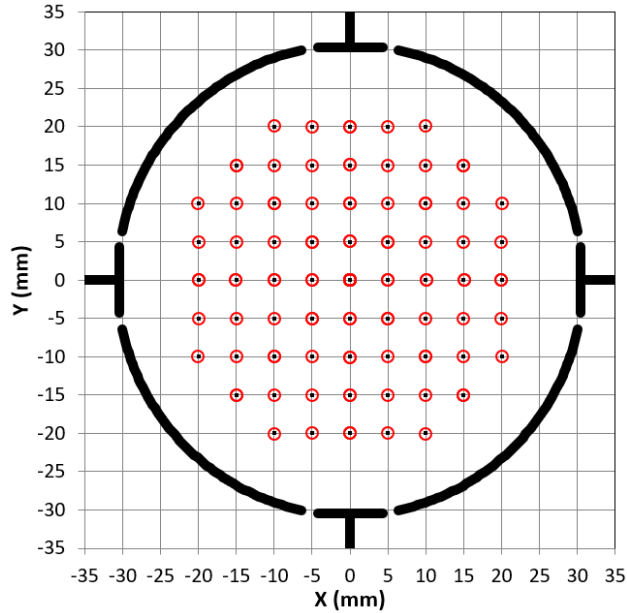

Figure 6 – A simple empirical correction has been applied to the positions shown in Fig. 4 (see text). The RMS distance between calculated (circles) and nominal positions (dots) is reduced from 230 μm to 29.2 μm, which makes position errors barely visible given the scales of this graph. The values used for the correction terms (see Eq. 17 and 18) are $\varepsilon = 0.0234$ and $b = -0.0144$.

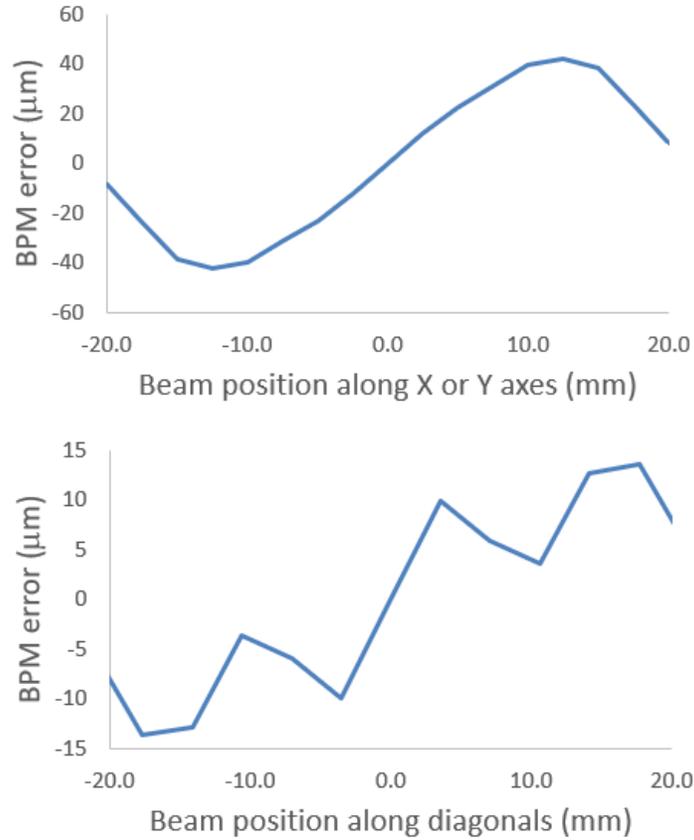

Figure 7 - Distances between the beam positions used in the simulations and positions calculated with the algorithm developed here using the simulation outputs plotted along the X and Y axes and along the diagonals. The difference compared to Fig. 5 is that an additional empirical correction has been applied, reducing the RMS error from 230 μm to 29.2 μm.

We compare the present results with results obtained with a commonly used system, such as the one implemented at the Relativistic Heavy Ion Collider (RHIC) [7], where the horizontal and vertical positions are obtained separately from the corresponding $Q_x$ and $Q_y$ values (see eq. 8) while improving the linear calibrations by adding cubic correction terms; i.e. terms proportional to $Q_x^3$ and to $Q_y^3$ respectively. The beam positions used in the simulation (black dots) and the positions calculated with these third order polynomials (red circles) are shown in Fig. 8.

We then, in Fig. 9, superimpose the position differences along a diagonal with the corresponding values obtain with the present approach, both with and without the correction terms. We see that, for this 60 mm diameter BPM with 10 mm diameter buttons, the present approach is more accurate for beam positions beyond ~3 mm from the center.

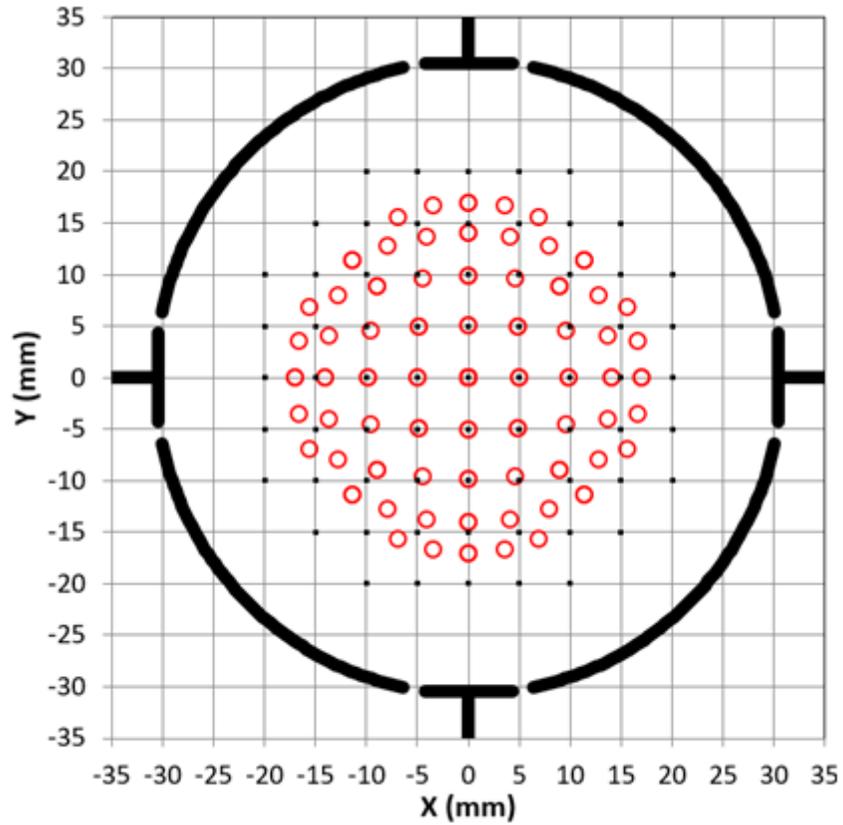

Figure 8 – The circles represent positions obtained by using third order polynomial calibrations applied individually to each axis[7], while the black dots represent actual beam positions. The large deviations for radii larger than ~10 mm can be contrasted with the much smaller deviations shown in figs. 4 and 6.

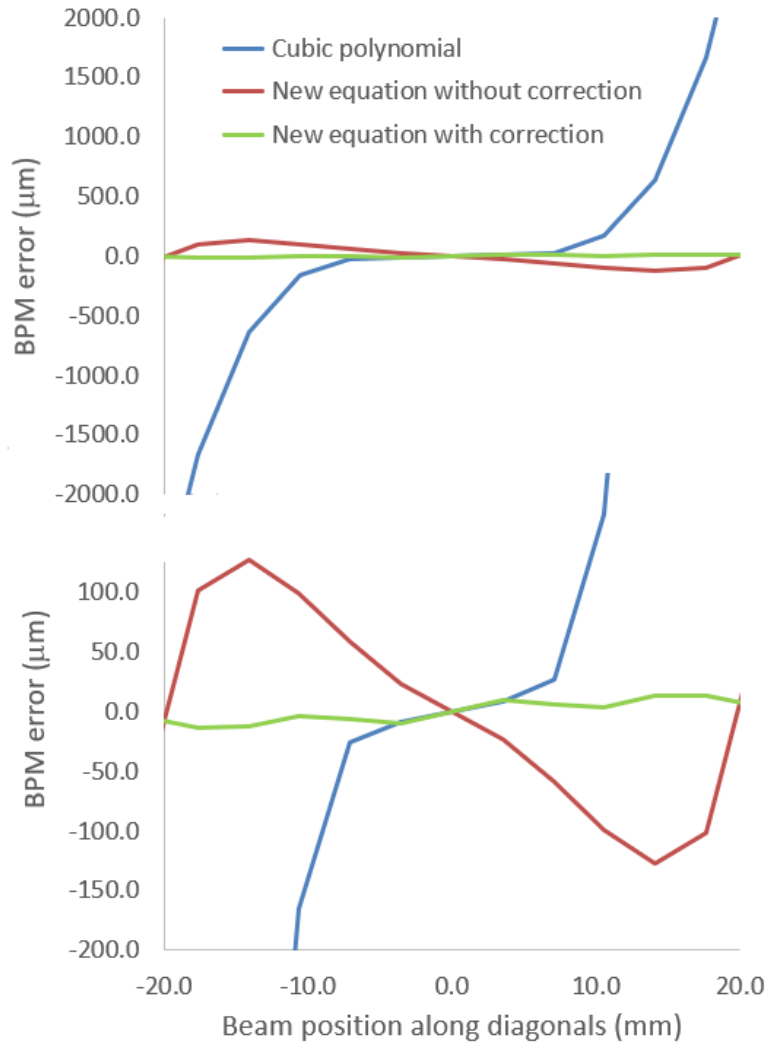

Figure 9 – BPM errors along the diagonal of the 60 mm diameter BPM computed with the conventional cubic polynomial approach and with the new equation with and without correction terms. The lower plot is a vertically expanded view of the upper one.

The results shown in Figs. 8 and 9 could in principle have been improved by adding cross-teems containing products of powers of $Q_x$ and $Q_y$. Such improvements, which would of course not have mitigated the discrepancies along the axes, would have been incompatible with the fast real-time response required, given the hardware that existed when the system was implemented. There is no reason to contemplate this type of improvement now, since much better results can be achieved with the present approach.

## V - Performance of the new algorithm with BPMs with finite width striplines

In this section we analyze the errors that occur if the simple analytic algorithm without and with corrections is used to determine beam positions in BPMs with increasingly wide striplines. Instead of generating simulated PUE signals with Particle Studio, as we did in the previous sections, we will now use the results of reference 1. Their equation 7 provides stripline signal amplitudes as function of beam position for striplines of any given width. To solve the inverse problem of finding beam positions, given PUE signal amplitudes, they use an iterative least squares procedure. The present algorithm derived for infinitesimally wide striplines is only approximately valid when the striplines are wider. We will explore here the magnitude of the deviations. For that purpose, we wrote a simple EXCEL VBA (Visual Basic for Applications) program that uses eq. 7 of reference 1 to calculate signal amplitudes for given beam positions and then, with these signals as inputs, uses the present algorithm to obtain calculated positions. The distances between the given and the calculated positions will then determine the performance of our algorithm for striplines of various widths. We used as our example a 100 mm diameter BPM, to make the results easily scalable to other diameters.

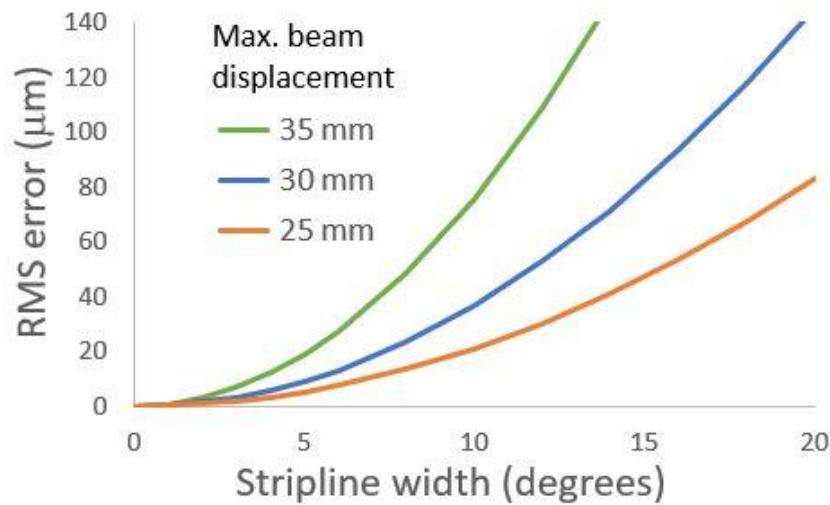

Figure 10 – RMS readout errors as function of stripline width over circular areas with radii that are 50%, 60% and 70% of the 50 mm radius BPM.

We see from Fig. 10 that striplines do not need to be of infinitesimal width to allow the use of the simple algorithm with errors that are quite small over a large fraction of the available maximum beam displacement. To see this more in detail, we plot in Fig. 11 readout errors along the horizontal and vertical axes and along the 45º diagonal for three of the stripline widths used in Fig. 10.

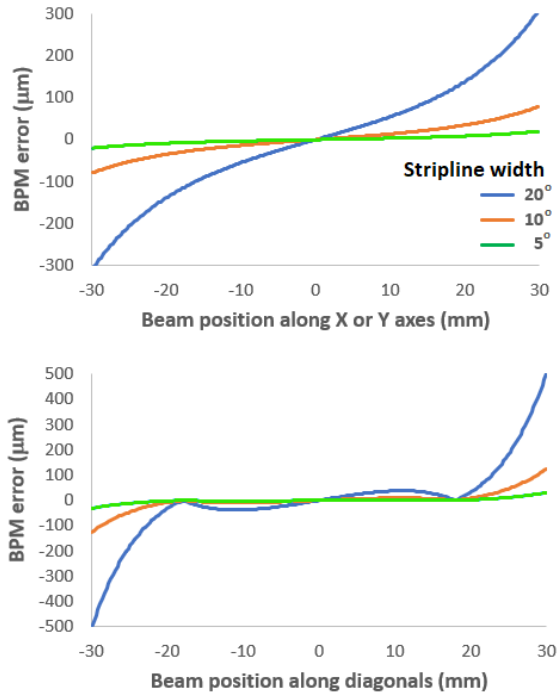

figure 11 – Readout errors along the x and y axes and along the diagonal in a 50 mm radius BPM when using the simple expressions derived in Section 1. Errors for striplines of three different widths are plotted.

Finally, we show in Fig 12, a 3-dimenional view of the deviations for the $10^0$-wide stripline case over one quadrant of the 100 mm diameter BPM. We see for example, both from Figs. 11 and 12 that errors are below 120 µm over a circular area with a radius that is 60% of the BPM radius.

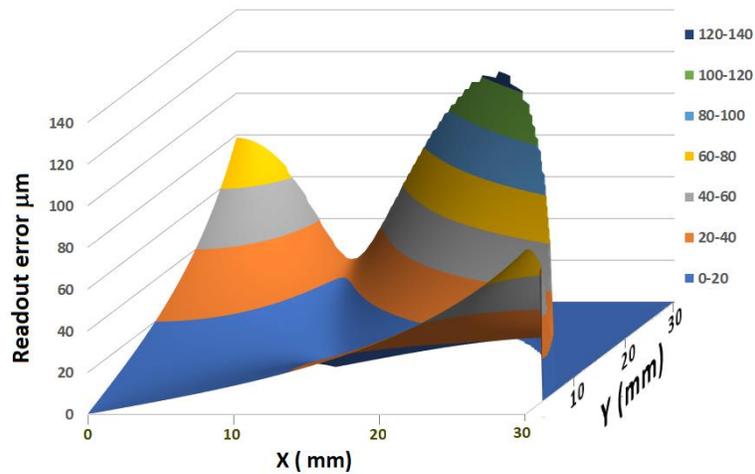

Figure 12 – Readout errors over one quadrant of a 100 mm diameter BPM with $10^0$ wide striplines

The simple correction terms described in section IV for the case of the button BPM studied in section III can also be applied here for the finite width stripline BPMs. In Fig. 13 we show examples calculated for a 50 mm radius BPM with 30° wide stripline PUEs. We see that very significant improvements are achieved by using the correction terms.

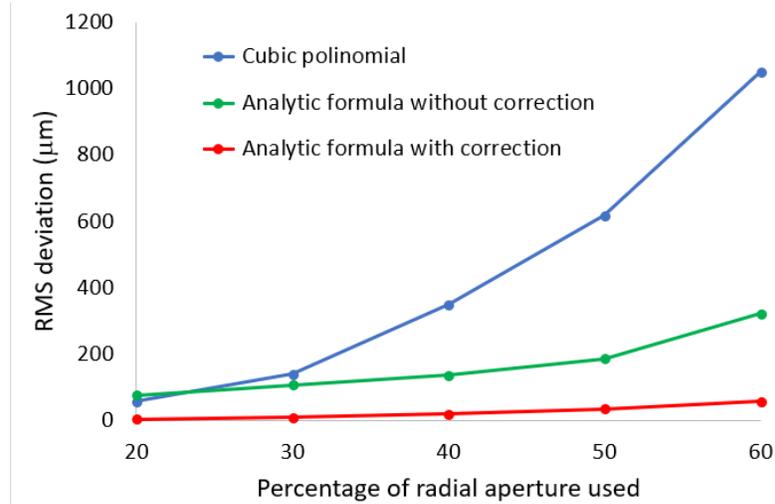

Figure 13 – Results calculated for a 50 mm radius BPM with 30° wide stripline PUEs. RMS deviations are shown as a function of the radii of circular areas for the standard cubic fit, for the new algorithm without correction and with correction. Corrections were optimized for each circular area. The largest values used for the correction terms (see Eq. 17 through 22) are $\varepsilon = 0.0225$ and $b = -0.0394$ for the last point at 60% of the BPM radius.

**VI – Performance for lower energy beams**

As mentioned in the introduction, equation 1, on which this algorithm is based, is only valid for relativistic beams, i.e. for beam velocity *v* close to the speed of light *c*. In this section we briefly explore the deviations that occur for values $\beta = v/c < 1$. For that purpose, we use Particle Studio simulations with the same 60 mm diameter BPM model used before and shown in Fig. 2. The values used for these simulations are 30 mm RMS Gaussian bunch length and $\beta$-values of 0.5, 0.7 and 0.9, corresponding to proton kinetic energies of 145.2 MeV, 375.6 MeV and 1214.3 MeV respectively. In addition, the $\beta = 1$ results are also included for comparison.

The results are shown in Fig. 14 for beam positions along the axes and along the diagonals. We plot the distances between the simulated beam positions and the positions calculated from the PUE signals that result from the simulations. The calculated positions are obtained with the present algorithm as defined by equations 17) through 22) with the usual definitions of $Q_x$ and $Q_y$ ratios given by equation 8). For each value of $\beta$, the correction parameters $b$ and $\varepsilon$ are chosen to minimize the errors

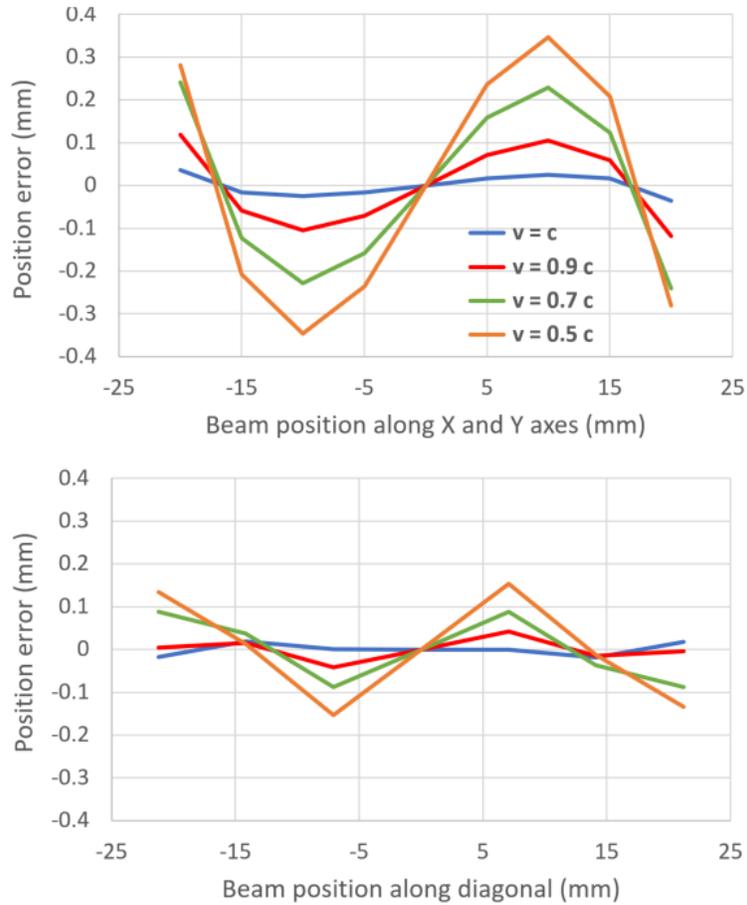

Figure 14 – Position errors as function of distance from the center for beams of different velocities obtained from simulations with the BPM model shown in Fig. 2 (see text).

The values of the parameters used as well as the maximum errors obtained are listed in Table 1

Table 1 - Simulation parameters and maximum errors.

| v/c | Proton energy (MeV) | $\varepsilon$ | b | Max. error (mm) |
|---|---|---|---|---|
| 1 |  | 0.022 | -0.0125 | 0.03 |
| 0.9 | 1214.3 | -0.0013 | -0.033 | 0.11 |
| 0.7 | 375.6 | -0.035 | -0.062 | 0.24 |
| 0.5 | 145.2 | -0.057 | -0.084 | 0.35 |

These simulations show increasing error values as the energy decreases. The beam position range chosen for these simulations is up to 20 mm from the center in this 30 mm radius BPM. For a smaller range, the optimum $\varepsilon$ and $b$ values for each energy would be different, and the maximum errors would be smaller.

For non-relativistic beams, the new algorithm provides good accuracy as well but with increasing corrections and larger error for the slower beams. In each case, the correction parameters will need to be determined through simulations or through measurements since they will not only be dependent on the value of *v/c* but also on the BPM geometry, the bunch length and the signal processing frequency range.

**VII - Implementations using Field Programmable Gate Arrays (FPGAs)**

Recently developed BPM electronics at BNL[7] utilize the Xilinx[8] 'Zynq' line of FPGAs that allows for very high speed floating-point calculations to be performed in hardware. A series of logic blocks provided by Xilinx have been used to create hardware that can perform basic mathematical operations, which are combined to calculate the beam position in real-time. An existing design that performs the 'difference over sum' method of position calculation (using a single pair of pickup electrodes) has been deployed in the electronics for some time[7]. Each math operation can be configured to take a variable amount of time (FPGA clock cycles) before providing a result. This setting directly affects how much FPGA resources each operation consumes. The routing of the signals in the FPGA is also made more complex with less clock cycles per operation, and at some point, the design becomes unworkable. The existing algorithm takes approximately 55 clock cycles to complete the position calculation. A clock rate of 200 MHz has been commonly used (5 ns period), yielding 275 ns of latency for the operation. Each individual part of the calculation is pipelined together, however, meaning that a new operand can be loaded into the beginning of the chain before the previous operand has completed computation. This allows for a calculation speed limited by the length of the longest part of the chain. The divide and square root functions are the most complex and have been set up to use 14 clock cycles to complete each result, which become the limiting elements. Therefore, a new position sample can be computed every 70 ns.

A new block of hardware, shown schematically in Fig.15, was added to the design to implement the aforementioned calculation method. The two 'ratio' terms $Q_x$ and $Q_y$ are taken from the existing single-plane calculation blocks, and act as the starting operands for the new four-pickup position calculation. These ratios are available after just 38 clock cycles. The limiting elements in the new formula are still divide, square-root, and (newly used) reciprocal blocks, using a 14 clock per operation setting, preserving the 70 ns position calculation rate. The total latency however has now increased, and with a length of 64 clocks adds another 320 ns to the result. These 64 clock cycles include the 52 cycles shown in Fig.17 and 12 additional cycles used to implement the empirical refinement discussed in section IV. Adding this on to the 38 clocks needed to get the 'ratio' terms to begin with, the latency is now 510 ns. The FPGA resources used for the four-pickup method were very close to what each of the original dual-pickup blocks consumed (~3000 lookup table or LUT resources), with the exception of using many more DSP (Digital Signal Processor) 'slices' (40 vs. 18, due to new reciprocal function). This additional usage is well within the headroom of our current FPGA capabilities and will allow this new algorithm to be tested on hardware with beam in a variety of accelerator applications in the near future.

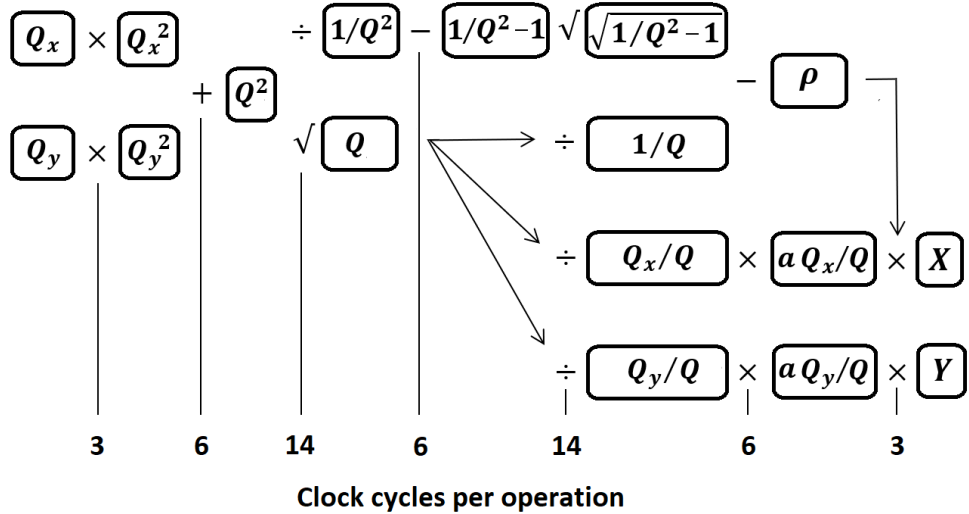

Figure 15 –Schematic diagram of the FPGA-based calculations used to implement the algorithm described in Section II. Each block is configured to perform a specific operation using the IEEE-754 single precision floating point representation using Xilinx[8] LogiCORE Floating-Point IP Blocks. The number of clocks to complete the longest operation is shown below each block of operations.

## VIII – Testing at the Cornell Photoinjector

In order to test how well this algorithm works in practice, we performed a brief test using the stripline BPMs (Fig. 16) at the Cornell Photoinjector [9]. The BPM was chosen because of its location at the end of a 1.5 m drift, after a pair of horizontal/vertical kicker magnets. The kicking magnets were slowly rastered in equal steps of magnet current over a grid, which was chosen such that the beam was nearly scraping the edges of the beam pipe at the BPM. All measurements were performed with ~5 pC bunch charge with a <1 µA of average current and a kinetic electron energy of 5.5MeV ($\beta = 0.9964$). The signal from the top, bottom, left, and right striplines were individually averaged and recorded for later analysis.

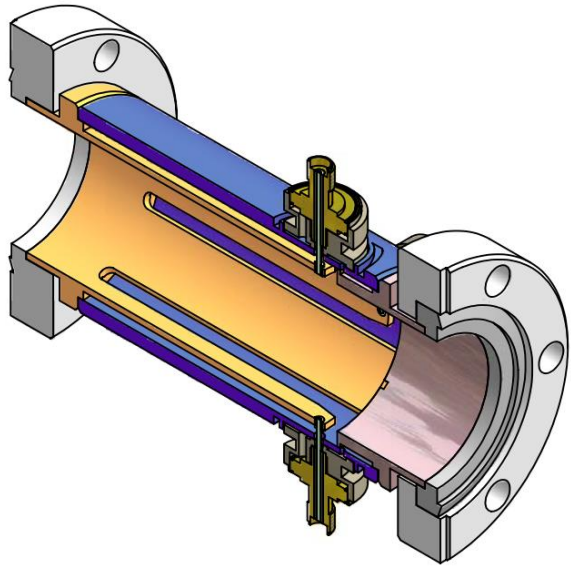

Figure 16 - Design of the stripline BPM used at the Cornell Photoinjector. The inner diameter of the pipe is 34.925 mm, and the striplines are 66 mm long, roughly 7.5 mm wide, and have 3.4 mm gaps on either side.

We analyzed the data by applying the simple difference/sum formula (Eqs. 15-16) and comparing to the corrected version of the present algorithm (Eqs. 17 through 22) for different values of the correction parameter *b*. For the purposes of this test, we kept $\varepsilon$ fixed at 0.0, as this does not affect the linearity of the resulting positions. As shown in Fig. 17, the difference/sum method produces positions only accurate within a few millimeters of the pipe center, while the nonlinearity-corrected algorithm can extend the valid range out to nearly the edge of the pipe. For this particular diameter of pipe and stripline width, a value of b = -0.08 seems to best correct the nonlinear curvature of the data.

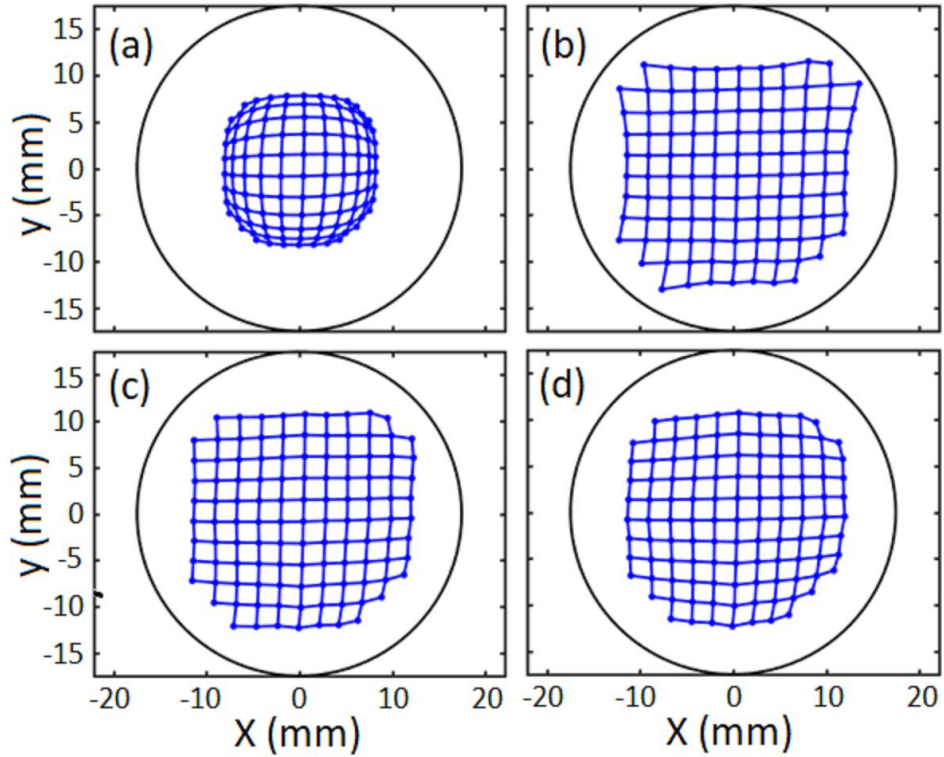

Figure 17 - Reconstructed beam positions using data measured with the BPM shown in Fig.16 for (a) the simple difference/sum method, and for the present algorithm with three different choices of the correction parameter b: (b) b = -0.06, (c) b = -0.08 and (d) b = -0.10. Here, points have been omitted if the signal intensity suggested that the beam was partially scraping the pipe.

## IX - Summary and conclusions

An analytic approach has been developed to calculate relativistic pencil beam positions in cylindrical BPMs with infinitesimal PUEs. It is shown that the normalized signal differences $Q_x$ and $Q_y$ can be considered as the components of a vector $Q$ which points in the direction of the beam. The position of the beam along the direction of $Q$ is a simple nonlinear function of the magnitude of $Q$. This position is then projected on the axes to obtain the coordinates of the beam.

We then analyzed the deviations that occur when applying this procedure to simulations with finite size PUEs and to cases where the beam isn't relativistic. The deviations found are surprisingly small. For the simulated button BPM and striplines of various widths, simple, empirically determined corrections reduced these errors even further.

The reduction of a two-dimensional problem to simple one-dimensional calculations has obvious computational advantages for the cases where the new algorithm is applicable. When cylindrical BPMs can be used with relatively small PUEs, corrections may not even be necessary. For applications with intense bunches, like those that are found in modern ion colliders and electron light sources, large PUEs are not necessary. They may in fact cause problems and limitations such as, for example, cryogenic BPM signal cable heating in RHIC [10].

The accurate position determinations for beams that are far from the center of the beam-pipe are of particular importance in cases where normal operation requires such orbits. That, for example, is the case for the CBETA project [11] that may serve as a recirculating electron Linac prototype for beam cooling in a future electron-ion collider [12]. The usual cubic approximation is totally inadequate in this case, even when the beam is in a plane defined by two of the PUEs.

The usual approach of approximating a non-linear response with a power series is far from optimal for BPM data. We discovered the analytic expression for ideal cases with v=c and small PUEs, and found that the Taylor expansion around the origin (Maclaurin expansion) of equation 12) converges very slowly. Therefore, many higher order terms beyond the third order would be required for good accuracy. We showed that the better approach is to use the analytic expression directly and to apply, simple, empirically determined corrections for the non-ideal cases.

Finally, an FPGA-based BPM readout implementation of the new algorithm was developed, allowing bunch intervals down to 70 ns with an output delay (latency) of only 510 ns. Tests with data from an actual BPM in the Cornell Preinjector[9] were successful.

The present approach offers significant accuracy and speed improvements for cylindrical BPM applications where possible beam offsets are sufficiently large to justify corrections to the linear approximation.

**Acknowledgement**

This work was supported by Brookhaven Science Associates, LLC, under Contract No. DE-AC02-98CH10886 with the US Department of Energy.**References**